\documentclass[preprint,aps,nofootinbib]{revtex4}
\usepackage{}
\usepackage{graphicx}%Include figure files
\usepackage{amsmath}
\usepackage{amsfonts}
\usepackage{amssymb}
\usepackage{color}%
\usepackage{dcolumn}% Align table columns on decimal point
\providecommand{\U}[1]{\protect\rule{.1in}{.1in}}

\definecolor{lightgray}{rgb}{.7,.7,.7}

\definecolor{red}{rgb}{1,0,0}

\definecolor{blue}{rgb}{0,0,1}

\definecolor{purple}{rgb}{0.6,0.1,0.7}

\newcommand{\f}{\begin{equation}}
\newcommand{\ff}{\end{equation}}
\newcommand{\fa}{\begin{eqnarray}}
\newcommand{\ffa}{\end{eqnarray}}

\begin{document}
\title{Fluid/Gravity duality with Petrov-like boundary condition in a spacetime with a cosmological constant }
\author{Tai-Zhuo Huang $^{1}$}
\email{huangtaizhuo@126.com}
\author{Yi Ling $^{2,1,5}$}
\email{yling@ncu.edu.cn}
\author{Wen-Jian Pan $^{1}$}
\email{wjpan_zhgkxy@163.com}
\author{Yu Tian $^{3,5}$}
\email{ytian@gucas.ac.cn}
\author{Xiao-Ning Wu $^{4,5}$} \email{wuxn@amss.ac.cn}

\affiliation{$^1$Center for Relativistic Astrophysics and High
Energy Physics, Department of Physics, Nanchang University,
330031, China\\ $^2$ Institute of High Energy Physics, Chinese
Academy of Sciences, Beijing 100049, China\\
$^3$ College of Physical Sciences, Graduate
University of Chinese Academy of Sciences, Beijing 100049, China\\
$^4$ Institute of Mathematics, Academy of Mathematics and System
Science, CAS, Beijing 100190, China and Hua Loo-Keng Key
Laboratory of Mathematics, CAS, Beijing 100190, China\\
$^5$ Kavli Institute for Theoretical Physics China(KITPC), Beijing
100080, China}

\begin{abstract}
Recently it has been shown that imposing Petrov-like condition on
the boundary may reduce the Einstein equation to the Navier-Stokes
equation in the non-relativistic and near-horizon limit. In this
paper we extend this framework to a spacetime with a cosmological
constant. By explicit construction we show that the Navier-Stokes
equation can be derived from both black brane background and
spatially curved spacetime. We also conjecture that imposing
Petrov-like condition on the boundary should be equivalent to the
conventional method using the hydrodynamical expansion of the metric
in the near horizon limit.

\end{abstract} \maketitle

\section {Introduction}
The hydrodynamical behavior of gravity has been greatly investigated
since Damour discovered that the excitations of a black hole horizon
behave very much like those of a fluid
\cite{Damour1979,P-T,Jacobson:1995ab,PSS,KSS,B-L,I-L,Bhattacharyya:2008kq,EFO,Padmanabhan10rp}.
In particular, recent progress on the Wilsonian approach to
Fluid/Gravity duality has opened a new window to link the Einstein
equation to the Navier-Stokes equation for a general class of
spacetime geometries \cite{Wilsonian,Heemskerk10hk,Faulkner10jy}. In
this setup the gravitational fluctuations are confined inside the
region between the horizon and a finite cutoff at radius $r=r_c$,
and a dual holographic fluid lives on the cutoff surface. The
hydrodynamical quantities can be obtained in the non-relativistic
limit through the standard procedure in AdS/CFT dictionary and their
dependence on the cutoff $r_c$ is interpreted as the renormalization
group flow in the fluid
\cite{NS-E,Compere11dx,Cai,Kuperstein,Brattan,Bredberg11xw,bbs,Matsuo11fk,
Bhattacharyya-Hubeny,Bhattacharyya:2008mz,Bhattacharyya:2008ji,Banerjee:2008th,Eling:2011ct}.

The holographic nature of the Petrov-like boundary condition has
been originally disclosed in \cite{Lysov11xx}. It has been shown
that embedding a hypersurface $\Sigma_c$ into a Minkowski spacetime,
a theory of gravity can be reduced to a theory of fluid without
gravity in one less dimension by simply imposing Petrov-like
condition on the boundary $\Sigma_c$. More explicitly, one finds
that in near horizon limit such kind of boundary conditions will
reduce the degrees of freedom of gravity such that the ``momentum
constraint'' gives rise to the incompressible Navier-Stokes
equation. In contrast to the conventional method using the
hydrodynamical expansion of the metric, imposing such kind of
boundary condition on the cutoff surface is much simpler and elegant
than putting regularity condition on the horizon. In this approach
the basic setup is to consider the perturbations of the extrinsic
curvature of $\Sigma_c$ while keeping the intrinsic metric fixed.
Furthermore, in this approach we treat the Brown-York stress tensor
as the fundamental variables, and identify its components with the
hydrodynamical variables of a fluid directly. One even need not to
know the explicit form of the perturbed metric in the bulk, thus no
need to solve the perturbation equations in the bulk either.
Recently, we have extended the framework above to a spacetime with a
spatially curved embedding in \cite{HL}.

In this paper we intend to further disclose the holographic nature
of Petrov-like boundary condition by extending the framework to a
spacetime with a cosmological constant. In this setting a black
brane background is allowed. It is very known that in this kind of
background the fluid/gravity duality has been extensively
investigated through the traditional hydrodynamic method with metric
expansion in the bulk and various hydrodynamic parameters have been
obtained for the dual fluid, see \cite{Cai} and references therein.
But it is completely unknown if imposing the Petrov-like condition
on the cutoff surface would lead to the same results in this
circumstance. We would like to stress that this is not a trivial
issue at all because no one can say that these two methods are
actually equivalent even in the near horizon limit. Technically, the
former (namely the hydrodynamic expansion) imposes the regularity
condition on the horizon and needs a long-wavelength expansion,
while the latter (the Petrov-like condition) imposes the boundary
condition on the cutoff surface and identifies the non-relativistic
limit with the near horizon limit. Therefore, when a cosmological
constant is taken into account, we are allowed to take the black
brane as the background and then compare our results obtained by
employing the Petrov-like conditions with the previous results
obtained through the hydrodynamic expansion of the metric in the
bulk. This is one of our main motivations for this paper. We will
firstly present the Petrov-like boundary condition in terms of the
Brown-York tensor and then consider the perturbations of the
extrinsic curvature while keeping the intrinsic metric fixed. The
Navier-Stokes equations with incompressibility conditions are
derived from black brane background and spatially curved spacetime,
respectively. In particular, for a black brane background we will
discuss the consistency of our results with those previously
obtained through the hydrodynamical expansion. The universality of
imposing Petrov-like boundary condition in a general spacetime and
its equivalence with the hydrodynamical expansion of the metric in
the near horizon limit are proposed in the end of this paper.

\section{the Framework for a spacetime with a cosmological constant}
For our purpose we assume the bulk metric in $p+2$ dimensional
spacetime has a general form as
\begin{equation}\label{ME}
{ds^2}_{p+2}=-f(r)dt^2+2d{t}dr+e^{\rho(r,x^i)}\delta_{ij}dx^idx^j,
\end{equation}
where $f$ is a function of radial coordinate $r$, while $\rho$
 depends on coordinates $r$ and $x^i$. First of all, we require that this metric
should be a solution to the vacuum Einstein equation with an
arbitrary cosmological constant $\Lambda$, namely
\begin{eqnarray}\label{EE}
G_{\mu\nu}=-{\Lambda}g_{\mu\nu},\ {\mu},{\nu}=0,{\ldots}, p+1.
\end{eqnarray}
Plugging the above metric into Eq.(\ref{EE}) leads to the following
equations
\begin{eqnarray}
   & &{\partial_r}{\partial_i}{\rho}                                                              =         0 \nonumber\\
   & &{\partial_r}^2{\rho}+{1\over{2}}({\partial_r}{\rho})^2                                      =         0 \nonumber\\
   & &{\partial_r}^2{f}+{p\over{2}}({\partial_r}f)({\partial_r}{\rho})                            =         -{4\Lambda\over{p}}
\label{comeq}\end{eqnarray} and
\begin{equation}
    {^{p+1}}R_{ij} =
    \gamma_{ij}[{1\over{2}}({\partial_r}f)({\partial_r}{\rho})+{pf\over{4}}({\partial_r}{\rho})^2
+{f\over{2}}{\partial_r}^{2}{\rho} + {2\Lambda\over{p}}].
\label{comeq2}\end{equation} It is straightforward to solve the
equations in (\ref{comeq}) and find the general solutions to be
\begin{eqnarray} \label{GS}
{f(r)}&=&-{2\Lambda\over{p(p+1)}}{(r+c)^2}-{{c_1}\over{(p-1)(r+c)^{p-1}}}+{c_2}\nonumber\\
{\rho (r,x^i)}&=&{F(x^i)}+{2\ln(r+c)},
\end{eqnarray}
where $c$, $c_1$ and $c_2$ are integration constants. Here we need
not to know the specific form of $F(x^i)$, which in principle can
be obtained by solving equation (\ref{comeq2}).\footnote{
We would like to claim that non-trivial solutions for the function
$F(x^i)$ always exist. According to Eqs. (\ref{comeq2}),
(\ref{GS}) and (\ref{Rij2}), we can obtain the constraint equation
of $F(x^i)$ as
$${c_2(p-1)\over{(r_c+c)^2}}\gamma_{ij} = {2-p\over{2}}\partial_i\partial_jF - {1\over{2}}\delta_{ij}\delta^{km}\partial_k\partial_mF
         + {p-2\over{4}}(\partial_iF)(\partial_jF)
                  -{p-2\over{4}}\delta_{ij}\delta^{km}(\partial_kF)(\partial_mF).$$
In section III, the black brane metric with $c_2=0$ and $F(x^i)=0$
satisfies the above equation automatically. Besides, we can always
construct non-trivial solutions for spatially curved spacetime as
discussed in section IV. For instance, the solutions could be
$F(x)=-2\ln{x^i}+\ln{{p(1-p)\over{2\Lambda+p}}}$, where $x^i$ can
be any spatial coordinate and $\Lambda<-{p\over{2}}$. As a matter
of fact, these solutions are nothing but the constant curvature
hyperbolic space $H^p$ with the Poincar\'e coordinates.} In next
two sections we will explicitly construct models with different
backgrounds by choosing appropriate values for these constants.

Now we consider an embedding $\Sigma_c$ with a
$p+1$ dimensional induced metric $\gamma_{ab}$ by setting $r=r_c$.
Since the spatial part of the hypersurface is conformally flat and
\begin{eqnarray} \label{Rij2}
  &  & ^{p+1}R_{ij} = {^p}R_{ij} =
  {c_2(p-1)\over{(r_c+c)^2}}\gamma_{ij},
\end{eqnarray}
one can show that the hypersurface must be a spacetime with a
constant curvature with the use of the Einstein equation. On this
cutoff surface its extrinsic curvature $K_{ab}$ should satisfy the
$p+1$ dimensional ``momentum constraint"
\begin{eqnarray}
D^a(K_{ab}-{\gamma_{ab}}K)=0,\label{mc}
\end{eqnarray}
as well as the ``Hamiltonian constraint"
\begin{eqnarray}\label{HC1}
{^{p+1}R}+{K_{ab}}{K^{ab}}-{K^2}-{2\Lambda} = 0,\label{hc}
\end{eqnarray}
where $D_a$ is compatible with the induced metric on $\Sigma_c$,
namely $D_{a}{\gamma_{bc}}=0$.

To impose Petrov-like boundary condition on this cutoff
surface\footnote{ Traditionally the Petrov conditions are used to
classify the geometry of spacetime by considering the multiplicity
of principle null vector at each point of spacetime. In fact, the
Petrov condition is a point-wise definition, so it can be
generalized to a hypersurface, i.e. only consider points on that
hypersurface. In this paper, Eq.(\ref{petrov}) is just the condition
we impose on the boundary, which is exactly the same as that in
\cite{Lysov11xx}. Nevertheless, to avoid possible confusion we would
like to rename it as Petrov-like boundary condition through this
paper.}, we decompose the $p+2$ dimensional Weyl tensor in terms of
$p+1$ dimensional quantities such that the Petrov-like condition can
be expressed in terms of the $p+1$ dimensional curvature and the
induced metric. Specifically, it can be decomposed as
\begin{eqnarray} \label{Weyl1}
    C_{abcd}      &=&      ^{p+1}R_{abcd}+ K_{ad}K_{bc}- K_{ac}K_{bd}  +
                                 {2\Lambda\over{p(p+1)}}(\gamma_{ad}\gamma_{bc}-\gamma_{ac}\gamma_{bd})
                                 \nonumber  \\
    C_{abc(n)}    &=&      D_a K_{bc} - D_b K_{ac}  \nonumber  \\
    C_{a(n)c(n)}  &=&      - \ ^{p+1}R_{ac} + K K_{ac} -
                                {K_a}^b K_{bc} +
                                {2\Lambda\over{p+1}}\gamma_{ac}, \label{Weyl}
\end{eqnarray}
where $\gamma_{ab}=g_{ab}-n_an_b$, $C_{abc(n)}=C_{abc\mu}n^{\mu}$,
and $n^{\mu}$ is the unit normal to $\Sigma_c$. The Petrov-like
 boundary condition on $\Sigma_c$ is defined as
\begin{eqnarray} \label{petrov}
C_{(\ell)i(\ell)j}=\ell^{\mu}{m_i}^{\nu}\ell^{\alpha}{m_j}^{\beta}C_{\mu\nu\alpha\beta}=0,
\end{eqnarray}
where $p+2$ Newman-Penrose-like vector fields satisfy the following
relations
\begin{eqnarray}
\ell^2=k^2=0,\ \ (k,\ell)=1,\ \ (k,m_i)=(\ell,m_i)=0,\ \
(m^i,m_j)={\delta^i}_j.
\end{eqnarray}
The reduction from the Einstein equation to Navier-Stokes equation
by imposing Petrov-like boundary condition can be understood by
counting the degrees of freedom on the cutoff surface. Taking the
extrinsic curvature as the fundamental variables, we have
$(p+1)(p+2)/2$ independent components, while the Petrov-like
boundary condition puts $p(p+1)/2-1$ constraints on the extrinsic
curvature, where we need subtract one simply as the Weyl tensor is
traceless. Therefore, the remaining degrees of freedom is $p+2$,
which may be identified as the energy density, pressure and the
velocity of the dual fluid living on the cutoff surface. In this
sense the relations between the pressure and the velocity of the
fluid is governed by the $p+1$ momentum constraints (\ref{mc}),
while the Hamiltonian constraint (\ref{hc}) is an equation
relating the energy density to the pressure of the fluid. In next
two sections we will explicitly demonstrate that $p+1$ momentum
constraints on the cutoff surface will give rise to the
incompressibility condition and the Navier-Stokes equation in the
near horizon limit.

\section{Navier-Stokes Equation From A Black Brane Background}
In this section we will employ the Petrov-like boundary condition
to a black brane background and then derive the Navier-Stokes
equation on the cutoff surface in the near horizon limit.
Previously this result has been obtained with a hydrodynamical
expansion of the metric in \cite{Cai}. Setting the integration
constants in Eq.(\ref{GS}) as
\begin{eqnarray}
   & &c_1=2m(p-1), \ \ \ c_2=0, \ \ \ c=0,  \nonumber \\
   & &\Lambda=-{p(p+1)\over{2}},  \ \ \ F(x^i)=0,
\end{eqnarray}
we can obtain a black brane metric as
\begin{equation}
    {ds^2}_{p+2} = -f(r)dt^2 + 2d{t}dr + r^2\delta_{ij}dx^idx^j,
\end{equation}
where $f(r)=r^2(1-{r_h^{p+1}\over{r^{p+1}}})$ and $r_h$ is the
position of the horizon. The hypersurface $\Sigma_c$ is located
outside the horizon at $r=r_c$, then the induced metric on the
hypersurface is
\begin{eqnarray}
    {ds^2}_{p+1}    &=&     -f(r_c)dt^2 + {r_c}^2\delta_{ij}dx^idx^j
                            \nonumber \\
                    & \equiv &     -{dx^0}^2 +
                    {r_c}^2\delta_{ij}dx^idx^j,
\end{eqnarray}
where we have defined $x^0=\sqrt{f(r_c)}t$. Obviously this is a
spatially flat embedding. In order to investigate the hydrodynamical
behavior of geometry in the non-relativistic limit, we further
introduce a parameter $\lambda$ by rescaling the time coordinate
with $\tau=\lambda x^0$  such that the metric has the form
\begin{equation}
    {ds^2}_{p+1}  =      -{1\over{\lambda^2}}d\tau^2 +
                            {r_c}^2\delta_{ij}dx^idx^j.
\end{equation}
The non-relativistic limit is achieved by setting
$\lambda\rightarrow 0$. In this coordinate system, the components
of the extrinsic curvature are
\begin{eqnarray}
    {K^\tau}_\tau &=& {1\over{2\sqrt{f}}}\partial_r f,\ \ \ \ \ \ {K^\tau}_i = 0, \nonumber\\
    {K^i}_j       &=& {1\over{r}}\sqrt{f}{\delta^i}_j,\ \ \ \ \ \ \,\ K  \ =  {1\over{2\sqrt{f}}}\partial_r f +
                                                                              {p\over{r}}\sqrt{f},
\end{eqnarray}
where $K$ is the trace of the extrinsic curvature. Employing the
Brown-York stress tensor which is defined on $\Sigma_c$ as
\begin{equation}
t_{ab}=K{\gamma_{ab}}-K_{ab},
\end{equation}
we can further express  the extrinsic curvature in terms of the
Brown-York tensor as follows
\begin{eqnarray} \label{EC}
    {K^\tau}_\tau &=&  {t_{tr}\over{p}}-{t^\tau}_\tau,\ \ \ \ \ \ \ \ \ \  \ \ \, {K^\tau}_i = -{t^\tau}_i, \nonumber\\
    {K^i}_j       &=&  -{t^i}_j+{\delta^i}_j{t_{tr}\over{p}},\ \ \ \ \ \ \ \  K \,\,\,=
    \,{t_{tr}\over{p}},
\end{eqnarray}
where $t_{tr}$ is the trace of $t_{ab}$. Now with the setup
presented above, we start to investigate the dynamical behavior of
the gravity on $\Sigma_c$ in the near horizon limit. First of all,
in contrast to the conventional perturbation method with metric
expansion, here we take the extrinsic curvature of the cutoff
surface as the fundamental variable while keeping the intrinsic
metric of the surface fixed. More conveniently, we may directly
consider the fluctuations of the Brown-York tensor on the surface,
and expand its components in powers of $\lambda$ as
\begin{eqnarray} \label{ttau}
    {t^\tau}_i       &=&    0 + \lambda{{t^\tau}_i}^{(1)} + \ldots  \nonumber\\
    {t^\tau}_\tau    &=&    {p\over{r}}\sqrt{f} + \lambda{{t^\tau}_\tau}^{(1)} + \ldots  \nonumber\\
    {t^i}_j          &=&    ({1\over{2\sqrt{f}}}\partial_r f + {p-1\over{r}}\sqrt{f}){\delta^i}_j
                            +  \lambda {{t^i}_j}^{(1)} + \ldots  \nonumber\\
    t_{tr}\,         &=&    ({p\over{2\sqrt{f}}}\partial_r f + {p^2\over{r}}\sqrt{f})
                            + \lambda {t_{tr}}^{(1)}+ \ldots
\end{eqnarray}
Now we concentrate on the perturbation behavior of gravity in the
near horizon limit. Mathematically the equivalence between the long
wavelength hydrodynamical expansion and the near horizon expansion
has been stressed in \cite{NS-E}, even at the nonlinear level. Here
we find the near horizon limit can be achieved simultaneously with
the non-relativistic limit through relating the perturbation
parameter $\lambda$ to $(r_c-r_h)$, namely, the coordinate distance
of the cutoff surface to the horizon. More explicitly, we find they
may be related by $\lambda^2=\alpha(r_c-r_h)$, with
$\alpha={1\over{(p+1)r_h}}$, which also implies that
$\tau=(r_c-r_h)t$.\footnote{The parameter $\alpha$ can be fixed as
follows. We firstly change the position of the horizon to
$r^\prime=0$ by a translation $r^\prime=r-r_h$, and then expand the
bulk metric of the black brane near the horizon, leading to
$ds^2=-(p+1)r_hr^\prime dt^2+2d{t}dr^\prime+r_h^2dx_idx^i$. Further
taking a coordinate transformation by $\tilde{t}=(p+1)r_ht$ and $
\tilde{r}={r^\prime\over{(p+1)r_h}}$, we find it is nothing but the
Rindler wedge of a Minkowski space. Since for the Rindler wedge, we
identify $\lambda^2$ with the location of the hyperbola by $
\lambda^2=\tilde{r_c} $, $\tau = \lambda^2 \tilde{t}$, for the black
brane it corresponds  to setting
$\lambda^2=\frac{r_c-r_h}{(p+1)r_h}$ in $(t,r)$ coordinate system .}
As a consequence, the near horizon expansion of the functions in
Brown-York tensor can be expressed in powers of $\lambda$ as follows
\begin{eqnarray}
{{\partial_r}f\over{\sqrt{f}}}|_{r_c}       &=&
                                                {1\over{\lambda}}-{3(p-2)(p+1)\over{4}}\lambda+{\ldots}  \nonumber\\
{\sqrt{f}\over{r}}|_{r_c}                   &=&     (1+p)\lambda-{(p+2)(p+1)^2\over{4}}\lambda^3+\ldots \nonumber\\
\sqrt{f}|_{r_c}                             &=&     r_h(p+1)\lambda
+
                                                    {r_h(p+1)\over{4}}(2+p-p^2)\lambda^3 +\ldots
\end{eqnarray}
Mathematically this identification leads to the mixing of the near
horizon expansion with the non-relativistic expansion in the
fluctuations. It is interesting enough to notice that this kind of
mixing plays an essential role in deriving the standard
Navier-Stokes equation with unit shear viscosity in this coordinate
system, and we will show this immediately. Firstly, we rewrite the
Hamiltonian constraint in terms of the Brown-York tensor as
\begin{equation}
     {({t^\tau}_\tau)}^2 - {2\over{\lambda^2}}\gamma^{ij}{t^\tau}_i{t^\tau}_j +
    {t^i}_j{t^j}_i - {(t_{tr})^2\over{p}} -2\Lambda = 0.
\end{equation}
Taking the expansion in powers of $\lambda$, we find the
non-trivial leading order of this equation gives rise to
\begin{equation}
    {{t^\tau}_\tau}^{(1)} =
    -2\gamma^{ij}{{t^\tau}}_i^{(1)}{{t^\tau}_j}^{(1)}.\label{ttt}
\end{equation}
By choosing the vector fields as
\begin{equation}
    {{\sqrt{2}}l}={\partial_{0}}-{n},\ \
    {{\sqrt{2}}k}=-{\partial_{0}}-{n},\ \
    {m_i}={1\over{r}}{\partial_i},
\end{equation}
similarly, with the use of Eq.(\ref{Weyl}) the Petrov-like condition
can be rewritten in terms of the Brown-York tensor as follows
\begin{equation}
    {t^\tau}_\tau {t^i}_j +
    {2\over{\lambda^2}}\gamma^{ik}{t^\tau}_k{t^\tau}_j -
    2\lambda{t^i}_{j,\tau}  - {t^i}_k{t^k}_j -
    {2\over{\lambda}}\gamma^{ik}{t^\tau}_{(k,j)} +
    {\delta^i}_j[{t_{tr}\over{p}}({t_{tr}\over{p}}-{t^\tau}_\tau)+{2\Lambda\over{p}}+2\lambda\partial_\tau{t_{tr}\over{p}}]
     = 0.
\end{equation}
Expanding the equation in powers of $\lambda$, we have %the leading term
\begin{equation}
    {{t^i}_j}^{(1)} = 2\gamma^{ik}{{t^\tau}_k}^{(1)}{{t^\tau}_j}^{(1)} - 2\gamma^{ik}{t^\tau}_{(k,j)}^{(1)}
                      + {\delta^i}_j{{t_{tr}}^{(1)}\over{p}}.\label{tij}
\end{equation}
The momentum constraint in flat cutoff surface has the form
\begin{equation}
    \partial_a {t^a}_b = 0.
\end{equation}
Plugging the solutions obtained in Eq.(\ref{tij}) into this equation
and identifying
\begin{eqnarray}
    {{t^\tau}_i}^{(1)} = {1\over{2}}\upsilon_i, \ \ \ & {t_{tr}}^{(1)} =
    {p\over{2}}P,
\end{eqnarray}
we can straightforwardly obtain the incompressibility condition and
the standard Navier-Stokes equation with unit shear viscosity as
\begin{eqnarray}
     \partial_i \upsilon^i = 0,                                                         & & \\
     \partial_\tau \upsilon_i + \upsilon^k \partial_k \upsilon_i -
     \partial^2 \upsilon_i + \partial_i P = 0.                                           & &
\end{eqnarray}
In the end of this section we remark that our results are consistent
with the previous results presented in Ref.\cite{Cai}, although the
fluid quantities are identified with gravity quantities in a
different manner and the Navier-Stokes equation looks not exactly
identical. As a matter of fact, the apparent discrepancy comes from
the use of different coordinate systems. Identifying the coordinates
$\tau$, $\tilde{\tau}$ and $x^i$ in \cite{Cai} with ours $t$, $x^0$
and $x^i$ respectively, the incompressible Navier-Stokes equation
obtained for arbitrary finite cutoff $r=r_c$ in \cite{Cai} becomes
\begin{equation}
    \partial_0\beta_i - \nu_c\partial^2\beta_i + \partial_iP_c +
    \beta^j\partial_j\beta_i = 0,
\end{equation}
%\begin{equation} \label{cc}
%    \tilde{\partial}_\tau\beta_i - \nu_c\tilde{\partial}^2\beta_i + \tilde{\partial}_iP_c +
 %   \beta^j\tilde{\partial}_j\beta_i = 0,
%\end{equation}
where the viscosity $\nu_c = {r_c\over{r_h(1+p)}} (1-
{r_h^{p+1}\over{r_c^{p+1}}})^{1/2}$. Apparently $\nu_c$ is
vanishing in the near horizon limit. However, if we transform the
coordinate system to $(\tau, x^i)$ by rescaling
\begin{equation}
    x^0 = {1\over{\lambda}}\tau,\label{rescale}
\end{equation}
the equation then becomes
\begin{equation} \label{NS1}
    \lambda\partial_\tau \beta_i - \nu_c\partial^2\beta_i + \partial_iP_c +
    \beta^j\partial_j\beta_i = 0.
\end{equation}
Furthermore, when the time is rescaled as Eq.(\ref{rescale}), by
definition the velocity and the pressure of the fluid should
correspondingly be rescaled as
\begin{eqnarray}
    \beta_i = \lambda \upsilon_i, \ \ \ P_c = \lambda^2
    P.\label{32}
\end{eqnarray}
Thus, Eq.(\ref{NS1}) changes into the form
\begin{equation}
    \partial_\tau \upsilon_i - {\nu_c\over{\lambda }}
    \partial^2\upsilon_i + \partial_iP +
    \upsilon^j \partial_j\upsilon_i = 0,
\end{equation}
where $\upsilon_i$ and $P$ are exactly the velocity and pressure of
the fluid as shown before. It is easy to see that in the limit $r_c
\rightarrow r_h$, ${\nu_c\over{\lambda}} = 1$ and one obtains the
standard incompressible Navier-Stokes equation with unit shear
viscosity, which is the same as what we have obtained. Finally, it
is worthy to point out that in \cite{Cai} this equation is obtained
for arbitrary finite cutoff but here we have only considered the
near horizon limit, thus one may also notice that in the
hydrodynamical expansion of the metric in the bulk, the process
depends on the number of spatial dimension $p$ and the Navier-Stokes
equation is derived only for some specific $p$, for instance $p=3$
\cite{Cai}. However, in our case since the near horizon limit is
taken, we find the result is general and independent of the spatial
dimension $p$.

\section{Navier-Stokes equation in spatially curved spacetime}
In this section we intend to derive the Navier-Stokes equations by
imposing Petrov-like boundary conditions in a spacetime with
non-vanishing Ricci tensor $^pR_{ij}$. First of all, for
convenience we prefer to fix the Rindler horizon at $r=0$. This
can be easily implemented by choosing the integration constants in
Eq.(\ref{GS}) as follows
\begin{eqnarray}
c=1,\ \ \ \  {c_1}={1}+{4\Lambda\over{p(p+1)}},\ \ \ \
{c_2}={2\Lambda\over{(p-1)p}}+{1\over{p-1}}.
\end{eqnarray}
With this choice we find the functions $f(r)$ and $\rho(r,x^i)$
can be expanded in the near horizon limit as
\begin{eqnarray}
{f(r)}&=&{r}-{({p\over{2}}+{2\Lambda\over{p}})}{r^2}+ {p^2+p+4\Lambda\over{6}}r^3+ {\ldots} \nonumber \\
{\rho(r,x^i)}&=&F(x^i)+2r-r^2+{2\over{3}}r^3+\ldots
\end{eqnarray}
Now we consider an embedded hypersurface located at $r=r_c$. The
induced metric on this surface is
\begin{eqnarray}
{{ds^2}_{p+1}}&=&-{f(r_c)}{dt^2}+{e^{\rho}}{\delta_{ij}}{dx^i}{dx^j}\nonumber\\
              &\equiv&-{dx^0}^2+{e^{\rho}}{\delta_{ij}}{dx^i}{dx^j}\nonumber\\
              &\equiv&-{{d\tau^2}\over{\lambda^2}}+{e^{\rho}}{\delta_{ij}}{dx^i}{dx^j},
\end{eqnarray}
where a parameter $\lambda$ is introduced to discuss the dynamical
behavior of gravity on $\Sigma_c$ in the non-relativistic limit.
Moreover, we identify the parameter $\lambda$ with the location of
hypersurface by setting $r_c={\lambda}^2$ such that a large mean
curvature can be satisfied in the near-horizon limit
$\lambda\rightarrow 0$. In coordinate system $(\tau, x^i)$, the
non-zero components of the connection are
\begin{eqnarray}
{\Gamma^k}_{i j}&=&{1\over{2}} ({\delta^k}_j \partial_{i} {\rho} +
{\delta^k}_i {\partial_j} {\rho}
 - {\delta^{km}} {\delta_{ij}} {\partial_m {\rho}}),
 \end{eqnarray}
which depend on the specific form of $F(x^i)$ and in general
describe a spatially curved spacetime. Now we turn to consider the
perturbations of gravity on this background. The extrinsic
curvature of the cutoff surface is
\begin{eqnarray}
{K^{\tau}}_i   &=&   0, \ \ \ \ \ \ \ \ \ \ \ \ \ \ \ \ \
{K^{\tau}}_{\tau}  =   {{\partial}_{r}{f}\over{2\sqrt{f}}}, \nonumber \\
{K^i}_{j}\, &=&
{1\over{2}}{\sqrt{f}}{\partial_r}{\rho}{\delta^i}_j,\ \ \ \ \  K =
{1\over{2\sqrt{f}}}\partial_r f +
{1\over{2}}p\sqrt{f}\partial_r\rho.
\end{eqnarray}
Similarly we consider the fluctuations of the Brown-York tensor
and expand it in powers of $\lambda$ as
\begin{eqnarray}
{t^{\tau}}_i&=&0+\lambda{{t^\tau}_i}^{(1)}+{\ldots} \nonumber \\
{t^{\tau}}_{\tau}&=&{{p}\over{2}}{\sqrt{f}}{\partial_r}{\rho}+{\lambda}{{t^\tau}_\tau}^{(1)}+{\ldots} \nonumber \\
{t^i}_j&=&({{{\partial_r}f}\over{2\sqrt{f}}}+{{p-1}\over{2}}{\sqrt{f}}{\partial_r}{\rho}){\delta^i}_j+\lambda{{t^i}_j}^{(1)}+\ldots \nonumber \\
{t_{tr}}&=&{({p\over{2\sqrt{f}}}{\partial_r}f+{{p^2}\over{2}}{\sqrt{f}}{\partial_r}{\rho})}+{\lambda}{t_{tr}}^{(1)}+{\ldots}
\end{eqnarray}
Moreover, since the location of the hypersurface $r_c$ is identified
with $\lambda^2$, we expand the following quantities in powers of
$r_c$
\begin{eqnarray}
{\partial_r{\rho}}|_{r_c}                    &=&     2-2{r_c}+2{r^2_c}+{\ldots}\nonumber\\
{\partial_r{f}}|_{r_c}                       &=&     1-(p+{4\Lambda\over{p}}){r_c}+{\ldots}\nonumber\\
f\ |_{r_c}                                   &=&     {r_c}-({{{p}\over{2}}+{2\Lambda\over{p}}}){r^{2}_{c}}+{\ldots}\nonumber\\
{\sqrt{f}}|_{r_c}
                                             &=&     {r_c}^{1/2}-({{{p}\over{4}}+{\Lambda\over{p}}}){r_c}^{3/2}+{\ldots} \nonumber\\
{\partial_r{f}\over{\sqrt{f}}}|_{r_c}        &=&     {r_c}^{-1/2}
-
                                                     {3(p^2+4\Lambda)\over{4p}}{r_c}^{1/2}
                                                     + \ldots
\end{eqnarray}
Next we turn to expand the Hamiltonian constraint as well as the
Petrov-like boundary conditions in powers of $\lambda$.
Substituting Eq.(\ref{EC}) into Eq.(\ref{HC1}), we find the
Hamiltonian constraint becomes
\begin{eqnarray}
{^{p+1}R}+{(t^{\tau}}_{\tau})^2-{2\over{\lambda^2}}{\gamma^{ij}}{t^\tau}_i{t^\tau}_j+{t^i}_j{t^j}_i-{(t_{tr})^2\over{p}}-{2\Lambda}
= 0.
\end{eqnarray}
It is easy to check that the background satisfies this condition
automatically at the order of ${\lambda^{-2}}$. While the
non-trivial sub-leading order is ${\lambda^0}$ which gives rise to
\begin{eqnarray}
{t^\tau}_{\tau}^{(1)}=-2{\gamma^{ij(0)}}{t^\tau}_i^{(1)}{t^\tau}_j^{(1)}.
\end{eqnarray}
Here we have used  ${^{p+1}R^{(0)}}=2\Lambda +p$ and expanded the
spatial metric
 $\gamma_{ij}$  as
\begin{equation}
     \gamma_{ij} = e^{F(x^i)}\delta_{ij}(1+r)^2\equiv \gamma^{(0)}_{ij}+r \gamma^{(1)}_{ij}+r^2
     \gamma^{(2)}_{ij}.
\end{equation}
Thus $\gamma^{(0)}_{ij}\equiv e^{F(x^i)}\delta_{ij}$. As we have
pointed out in \cite{HL}, the ``spatially covariant derivative''
$D_i$ compatible with $\gamma_{ij}$ is also compatible with
$\gamma^{(n)}_{ij}$ since the connection is $r$-independent. We
will use this fact when deriving the Navier-Stokes equation from
the momentum constraint. Now we turn to the Petrov-like boundary
condition. We choose the vector fields as
\begin{equation}
{{\sqrt{2}}l}={\partial_{0}}-{n},\ \
{{\sqrt{2}}k}=-{\partial_{0}}-{n},\ \
{m_i}=e^{-{\rho}\over{2}}{\partial_i},
\end{equation}
then with the use of Eq.(\ref{Weyl1}), we find the boundary
condition Eq.(\ref{petrov}) can be finally written in terms of the
Brown-York stress tensor as well as the intrinsic curvature as
\begin{eqnarray}
 &&     {t^\tau}_{\tau}{t^k}_{j}+{\delta^k}_{j}{[{t_{tr}\over{p}}{({t_{tr}\over{p}}-{t^\tau}_{\tau})}+{2\lambda\over{p}}{\partial_{\tau}t_{tr}}]}
                    +{2\over{\lambda^2}}{\gamma^{ki}}{t^\tau}_i{t^\tau}_j-{2\lambda}{t^k}_{{j},{\tau}}\nonumber\\
 &&      -{2\gamma^{ki}\over{\lambda}}{t^{\tau}}_{(i,j)}
                    -{t^k}_{m}{t^m}_{j}-{\gamma^{ki}}{R_{ij}}+{2\gamma^{ki}\over{\lambda}}{\Gamma^m}_{ij}{t^\tau}_m+{2\Lambda\over{p}}{\delta^k}_j=0.
\end{eqnarray}
Similarly it is easy to check that the background satisfies this
condition automatically at the order of ${\lambda^{-2}}$, while at
the order of ${\lambda^0}$ we obtain the following result
\begin{eqnarray} \label{sl2}
    {{t^k}_j}^{(1)}={2\gamma^{ki(0)}}{{t^\tau}_i}^{(1)}{{t^\tau}_j}^{(1)}-{2\gamma^{ki(0)}}{t^\tau}_{(i,j)}^{(1)}
    +{{t_{tr}}^{(1)}\over{p}}{\delta^k}_j+{2\gamma^{ki(0)}}{\Gamma^m}_{ij}{{t^\tau}_m}^{(1)}.
\end{eqnarray}
Next we put this solution into the momentum constraint to reduce the
degrees of freedom of gravity. Now the momentum constraint on
$\Sigma_c$ is
\begin{eqnarray}
{D_a}{t^a}_b=0.
\end{eqnarray}
The time component of this equation at leading order gives rise to
\begin{eqnarray}
{D_i}{t^{\tau i(1)}}=0,
\end{eqnarray}
while the spatial components can be written as
\begin{eqnarray}
{\partial_{\tau}}{t^\tau}_i^{(1)}+{D_k}{t^k}_i^{(1)}=0.
\end{eqnarray}
Plugging the solution in Eq.(\ref{sl2}) into this equation, we
have
\begin{equation}
    \partial_\tau{{t^\tau}_i}^{(1)} + 2t^{\tau k
    (1)}D_k{{t^\tau}_i}^{(1)}+{1\over{p}}D_i{t_{tr}}^{(1)} -
    (D^kD_k{{t^\tau}_i}^{(1)}+{{t^\tau}_m}^{(1)}{R^m}_i) = 0.
\end{equation}
Similarly, we identify the remaining Brown-York variables with the
hydrodynamical variables as
\begin{eqnarray}
    {t_{tr}}^{(1)}={p\over{2}}P, \ \ \
    \ {{t^\tau}_i}^{(1)}={\upsilon_i\over{2}},
\end{eqnarray}
then the incompressibility condition and Navier-Stokes equation in
spatially curved spacetime can be obtained as follows
\begin{equation}
    D_i\upsilon^i = 0,
\end{equation}
\begin{equation}
    \partial_\tau \upsilon_i + \upsilon^kD_k\upsilon_i + D_iP -
    (D^kD_k\upsilon_i+{R^m}_i\upsilon_m)  = 0.
\end{equation}

\section{Summary and Discussions}
In this paper we have generalized the framework presented in
\cite{Lysov11xx,HL} to a spacetime with a cosmological constant. We
have demonstrated that the incompressible Navier-Stokes equation can
be derived from the Einstein equation by simply imposing the
Petrov-like boundary condition in the near horizon limit such that
the holographic nature of the Petrov-like boundary condition has
been further disclosed. Furthermore, we have shown that our results
are consistent with the ones previously obtained by hydrodynamic
expansions for a black brane background. We remark that the
fundamental variables such as the velocity and the pressure for a
fluid are introduced in different manners for these two methods. For
hydrodynamic expansion the velocity of the fluid is identified with
the fluctuations of the metric, while for Petrov-like boundary
condition it is identified with the component of the Brown-York
tensor. One may wonder why these different identifications do give
rise to the same dynamical equations for a fluid. For a black brane
background we have presented a detailed comparison between the
method of Petrov-like boundary condition and the method of the
hydrodynamic expansion of the metric in the near horizon limit.
After having figured out the connections of those hydrodynamic
quantities which are identified with gravity quantities in different
manners in these two methods, we find the values of the shear
viscosity obtained through these two methods are exactly identical
such that the reliability of the gravity/fluid duality has been
testified in a more solid foundation. On the other hand, this
consistency indicates that Petrov-like boundary condition contains a
holographic nature indeed, with some universality in linking the
Einstein equation to the Navier-Stokes equation. Our observation
here is crucial for us to better understand these two methods and
finally to be able to prove the conjecture that they may be
equivalent in the near horizon limit. The next step is to testify
its validity in a more general setting. First of all, we have
obtained the Navier-Stokes equation on a spatially curved
hypersurface, but only for a class of spacetime with constant
curvature, we expect this framework can be applied to a background
with a more general metric than what we have proposed in
Eq.(\ref{ME}). Secondly, when a matter field is taken into account,
besides extending the framework to including the contribution from
the matter field in the Petrov-like boundary condition, it also
involves how to put appropriate boundary conditions on the cutoff
surface for the matter field itself when it is also dynamical and
has some degrees of freedom. We will construct a model with Maxwell
field and provide an affirmative answer to this issue elsewhere
\cite{petrov3}. Therefore, based on all investigations mentioned
above we may conjecture that at least in the near horizon limit,
imposing Petrov-like condition on the boundary should be a universal
method to reduce the Einstein equation to the Navier-Stokes equation
for a general spacetime in the presence of a horizon. Furthermore,
in such a limit it may be equivalent to the conventional method
using the hydrodynamical expansion of the metric, where the
perturbation equations in the bulk are explicitly studied and the
regularity condition is imposed on the horizon. We leave this open
issue for further investigation in future.

Since the higher order hydrodynamical expansion has been
considered in \cite{Compere11dx}, and the higher order corrections
to the Navier-Stokes equations have also been derived there, we
think it is an interesting issue to study the higher order
expansions of the boundary conditions and constraints in the
approach of imposing Petrov-like boundary conditions. Such
investigations would provide us more understandings on correction
terms in the Navier-Stokes equation for real fluids.

In this approach the mixing of the perturbation expansion and the
near horizon expansion play a very interesting role in obtaining
the desired results. On the other hand, as we know the
Navier-Stokes equation can be derived on arbitrary finite cutoff
surface with the method of long wavelength hydrodynamical
expansion, and this approach is mathematically equivalent to the
near horizon expansion even at the nonlinear level \cite{NS-E}.
Thus we are wondering if the method of imposing Petrov-like
boundary condition can be applicable to the regime beyond the near
horizon limit, and our investigation is under progress.

\begin{acknowledgments}
Y. Ling, Y. Tian and X. Wu would like to thank the KITPC for
hospitality during the course of the programm ``String
Phenomenology and Cosmology '' when this work is  completed. T.
Huang, Y. Ling and W. Pan are partly supported by NSFC
(10875057,11178002), Fok Ying Tung Education Foundation
(No.111008), the key project of Chinese Ministry of Education
(No.208072), Jiangxi young scientists (JingGang Star) program and
555 talent project of Jiangxi Province. Y. Tian and X. Wu are
partly supported by NSFC (Nos. 10705048, 10731080 and 11075206)
and the President Fund of GUCAS. We also acknowledge the support
by the Program for Innovative Research Team of Nanchang
University.
\end{acknowledgments}

\end{document}